\documentclass[aip, pof,graphicx]{revtex4-2}

\usepackage{graphicx}
\usepackage{dcolumn}
\usepackage{bm}
\usepackage[utf8]{inputenc}
\usepackage[T1]{fontenc}
\usepackage{color}
\usepackage{longtable}
\usepackage{amssymb}
\usepackage{epstopdf}

\draft 

\begin{document}


\title[Bidisperse beds sheared by viscous fluids: Grain segregation and bed hardening]{Bidisperse beds sheared by viscous fluids: Grain segregation and bed hardening\\
	\textcolor{blue}{This article may be downloaded for personal use only. Any other use requires prior permission of the author and AIP Publishing. This article appeared in Phys. Fluids 35, 103326 (2023) and may be found at https://doi.org/10.1063/5.0168415.}}



\author{Jaime O. Gonzalez}
\altaffiliation[Also at ]{School of Mechanical Engineering, UNICAMP - University of Campinas, Campinas--SP, Brazil}
\affiliation{Departamento de Petróleos, Escuela Politécnica Nacional,\\
	Av. Ladrón de Guevara E11-253, Quito, Ecuador
}%

\author{Fernando D. C\'u\~nez}
\affiliation{Department of Earth and Environmental Sciences, University of Rochester,\\
	Rochester, NY 14627, USA}

\author{Erick M. Franklin*}%
 \email{erick.franklin@unicamp.br}
 \thanks{*Corresponding author}
\affiliation{School of Mechanical Engineering, UNICAMP - University of Campinas,\\
Rua Mendeleyev, 200, Campinas, SP, Brazil
}%


\date{\today}

\begin{abstract}
When a granular bed is sheared by a fluid that flows above a critical limit, it undergoes a complex motion that varies along time: it can contain fluid- (bedload) and solid-like (creep) regions, being prone to strain hardening and, in case of polydispersity, segregation. In this paper, we investigate experimentally the short- and long-time evolution of a bidisperse bed sheared by a viscous liquid. Different from previous experiments, the density ratio between grains and fluid is 2.7, close to values found in rivers and oceans. We show the existence of diffusive, advective and constrained regions, that most of segregation occurs during the very first stages of the flow, and that bed hardening becomes stronger while bedload and creep weaken along time. We obtain the segregation rates, their evolution along time, their variation with the applied shearing, and the time evolution of creeping and bedload. Finally, we propose characteristic times for the segregation of large particles and bed hardening. Our results shed light on the complex motion of sheared beds existing in nature, such as river beds and creeping lands.
\end{abstract}


\maketitle 

\section{INTRODUCTION}
\label{sec:intro}

The transport of grains by a fluid flow is frequently observed in nature, such as can be found in rivers and deserts. Whenever the ratio between the entraining (due to the fluid shearing) and resisting (due to gravity) forces is within moderate values, bedload and creep can occur within the granular bed. In the case of air, bedload consists of saltating grains that effectuate ballistic flights and which, by impacting onto the bed, move part of the non-saltating grains by creep motion \cite{Bagnold_1, Raudkivi_1}. In the case of liquids, bedload is a moving layer in which grains roll, slide or effectuate small jumps while keeping contact with the lower part of the bed. This lower part has been described as being static \cite{Raudkivi_1, Yalin_1}, but recently it has been shown that it may creep, with movements caused by very slow rearrangements of grains \cite{Houssais_1}. In addition, Houssais et al. \cite{Houssais_1} and Allen and Kudrolli \cite{Allen_2} showed that the creeping layer can exist even when shear stresses are below the threshold for bedload (so that a bedload layer is absent). In particular, Houssais et al. \cite{Houssais_1} showed that within the granular bed there is a continuous transition between bedload and creep, and proposed that this transition occurs at a height characterized by a viscous number \cite{Jop} $I_v$ equal to 10$^{-7}$, where $I_v$ is the ratio between the microscopic (related to the rearrangements of grains) and macroscopic (related to the macroscopic rate of deformation) timescales.

Sheared granular beds usually experience hardening, with grains having their mobility reduced along time. Bed hardening has been identified as one of the causes of the increase along time of the bedload threshold, the minimum shear stress necessary for bedload to take place. With flume experiments, Charru et al. \cite{Charru_1} and Masteller and Finnegan \cite{Masteller_1} measured the decay in the mobility of a bedload layer and proposed that the decrease was due to bed hardening, in its turn caused by purely geometric rearrangements of grains, i.e., the simple percolation of grains migrating to vacancies (leading to an increase in bed compactness). This explanation implies that bed hardening would be of isotropic nature. Later, Masteller et al. \cite{Masteller_2} identified hardening of a river bed by analyzing a 19-year-series dataset of fluid stress and sediment transport measured in the Erlenbach river.

Over the past decades, many authors have investigated the jamming of granular materials under normal and shear stresses \cite{Cates, Majmudar, Bi}, which bears some connection with the bed hardening observed in granular beds sheared by fluids. Cates et al. \cite{Cates} showed that fragile states may appear in colloidal suspensions and granular materials by the formation of force chains aligned in preferential directions, the materials being able to support loading, and then jamming, in such directions, but being unable to support loading in other ones, with consequent unjamming. Bi et al. \cite{Bi} showed that granular matter is subject to fragile states and shear jamming when external shear stresses are applied, in addition to the isotropic jamming that appears even in shear-free conditions. They observed that both fragile and shear-jammed states appear at lower particle fractions than those necessary for isotropic jamming, the fragile state appearing under small shear stresses and being characterized by a one-directional force network, while the shear-jammed state appears under stronger shear stresses and is characterized by a force network that percolates in different directions. The appearance of these different states may be regarded as memory formation, where out-of-equilibrium systems may keep information about the past \cite{Keim}. 

Recently, C\'u\~nez et al. \cite{Cunez2} carried out experiments in a circular channel to investigate the response of a granular bed to fluid-shear stress cycles of varying magnitude and direction, and determined the isotropic (due to bed compaction) and anisotropic (due to shear-induced orientation) contributions. They showed that the application of an external shearing in a given direction produces, along time, an anisotropic structure that keeps memory of the applied shear, with the corresponding fragile and/or jamming states. When, however, the shear direction is reversed, the former anisotropic structure is wiped out, causing memory erasure and the formation of a new anisotropic structure after a characteristic time. They found that sediment transport promotes direction-dependent strain hardening for moderate shear stresses, due to an accumulated memory from the past, while higher stresses fluidize part of the bed, engendering dilation-induced weakening and memory loss. Finally, they quantified the hysteresis in sediment transport depending on the orientation of varying flows.

In addition to hardening caused by bed compactness and shear-induced orientation, polydisperse beds may have their mobility reduced by natural armoring: the segregation leading to a higher concentration of larger grains on the bed surface. Those grains shield smaller particles from regions where the flow is stronger, hardening the bed \cite{Frey}. Ferdowsi et al. \cite{Ferdowsi} carried out experiments where a bidisperse granular bed was sheared by a steady viscous flow, and numerical simulations of a bed sheared by a layer of particles (without fluid). In their experiments, however, the ratio between the solid ($\rho_s$) and fluid ($\rho$) densities was $S$ = 1.13 (close to unity), so that the gravity effects were much lower than in natural flows of water and sand ($S$ $\approx$ 2.65). They found that bed armoring is mainly due to segregation, with the upward motion of large particles occurring from lower regions in the bed. They also showed the existence of two distinct layers: one, close to the surface, where bedload propel a fast shear-dependent segregation and which they associate with an advection mechanism, and another one below, where creep drives a slow segregation and which they associate with a diffusion-like mechanism.

Segregation in bidisperse bedload and debris flow were also investigated at relatively low timescales (those for which creeping cannot be measured). Zhou et al. \cite{Zhou2} investigated the effect of the interstitial fluid in particle segregation taking place in debris flows. By comparing the outputs of numerical simulations with and without the presence of an interstitial fluid (for both inertial and viscous fluids), they found that segregation is weaker and slower in the presence of an interstitial fluid, and that it weakens with increasing the fluid viscosity. The authors propose that both buoyancy and shear-rate alterations by the presence of fluids change the particle-particle dynamics that leads to the upward motion of large particles. Those results were later corroborated by Cui et al. \cite{Cui}, who investigated numerically the effect of different interstitial fluids in a confined granular medium under imposed shearing. The authors found that, indeed, the segregation decreases with increasing the fluid viscosity above a lower limit. Below that limit, segregation does not vary with viscosity, but with the density ratio and fluid inertia. Cui et al. \cite{Cui} propose that viscous effects weaken particle-particle contacts and dampen particle fluctuations, decreasing the rate of segregation. Rousseau et al. \cite{Rousseau} inquired into the upward motion of a single large particle (intruder) within a bedload layer of smaller particles. They found that the upward motion of the intruder has two phases, a first one which is intermittent and slow, and a second one consisting of a fast motion to the bed surface. However, it is possible that the first phase occurs in the limit between the creep and bedload layers (not in the bedload layer itself). In a different approach, Frey et al. \cite{Frey2} carried out experiments in which they tracked a sinking layer of smaller particles within a bidisperse bedload layer under turbulent flow. The authors found that the sink velocity of small particles decreases logarithmic in time and varies with the particle-particle shear rate, which decays exponentially with depth. They also found that small particles reach a final depth, forming a layer there. We conjecture that, perhaps, that layer takes place in the limit between the bedload and creep layers. 

Although previous works increased our knowledge on the importance of shear and segregation for bed armoring, with the consequent variation in sediment transport, questions such as the structure of shear-induced hardening in bidisperse beds, segregation rates, and long-time evolution for moderate-weight beds ($S$ > $1.5$) remain open. In this paper, we investigate the evolution of a bidisperse bed sheared by a viscous liquid. For that, we carried out experiments in which we made use of RIM (refractive index matching) visualizations and, different from previous experiments, the ratio between grains and fluid was 2.7, close to values found in rivers and oceans. We show the existence of diffusive, advective and constrained regions, that most of segregation occurs during the very first stages of the flow (first 20--80 minutes), and that bed hardening becomes stronger while bedload and creep weaken along time. We obtain the segregation rates, their evolution along time, their variation with the applied shearing, and the time evolution of creeping and bedload. Finally, we propose characteristic times for both the segregation of large particles and bed hardening. Our results provide new insights into the physical mechanisms of segregation and bed hardening occurring in nature, such as in river beds and creeping lands. In particular, we show how segregation takes place in polydisperse beds found in nature (leading to bed armoring), how the lower layer (creeping layer) of a granular bed compacts (promoting bed hardening), and how grains are rearranged within the bed (which hardens the bed while keeping memory effect \cite{Cunez2}). These results are important for understanding sediment transport found in  geophysical flows, hydraulics, and engineering applications.

In the following, Sec. \ref{sec:experiments} presents the experimental setup, Sec. \ref{sec:results} shows the results and Sec. \ref{sec:conclusions} concludes the paper.

\section{EXPERIMENTAL SETUP}
\label{sec:experiments}

\begin{figure}[h!]
	\begin{center}
		\includegraphics[width=.95\linewidth]{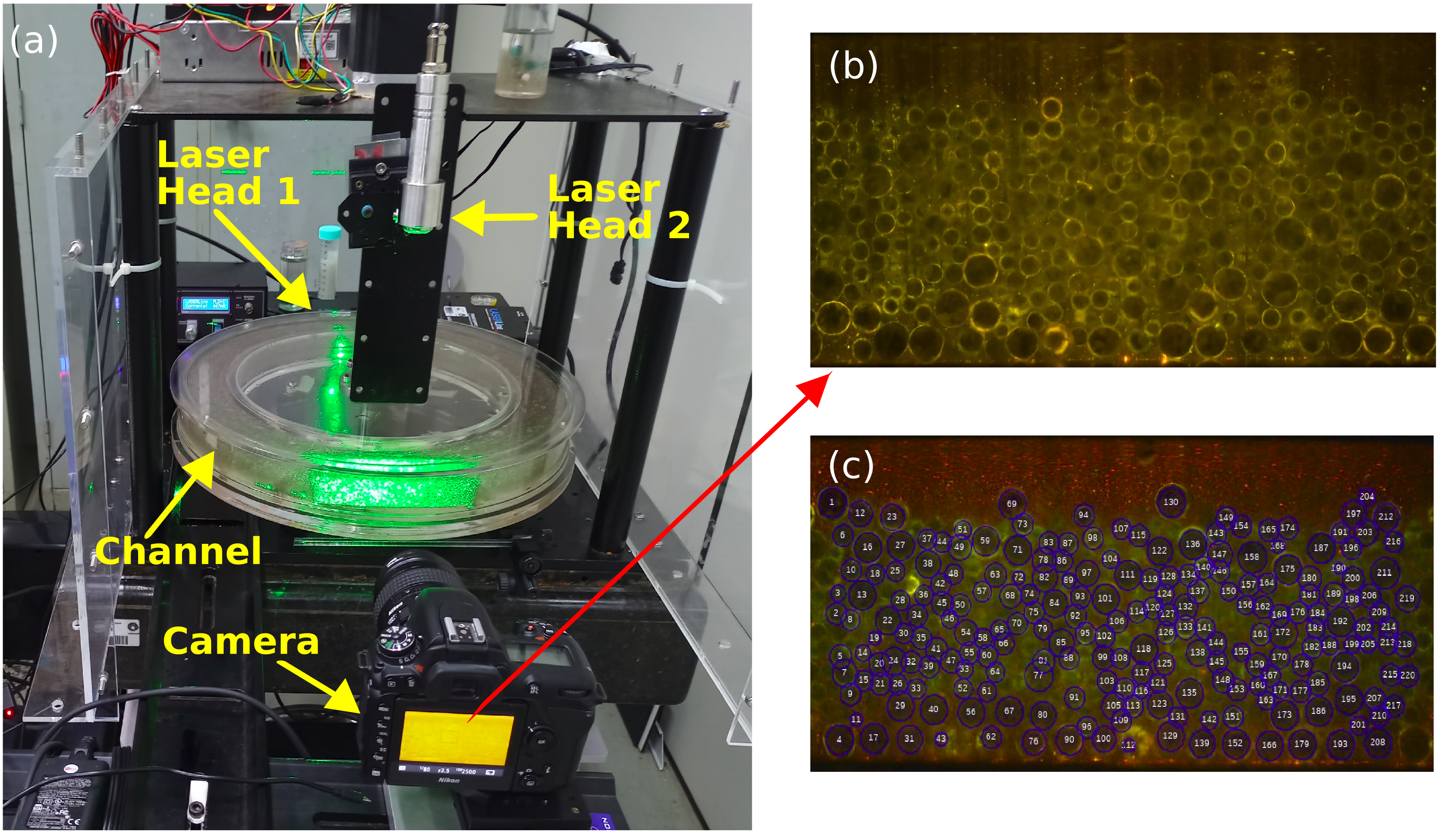}\\
	\end{center}
	\caption{(a) Photograph of experimental setup; (b) example of raw image; (c) detection of grains superimposed with a raw image.}
	\label{fig:setup}
\end{figure}

The experimental device consisted basically of an annular (circular) flume with a rotating lid connected to a computer-controlled stepper motor, so that the lid rotation imposed a shear driven flow inside the flume. The flume had mean radius $R$ = 18 cm, internal width $W$ = 40 mm, and internal height $H$ = 30 mm, being completely filled with controlled grains and liquid (described next), and the ensemble was mounted over an optical (heavy) table aligned horizontally. Figure \ref{fig:setup}(a) shows a photograph of the ensemble mounted over the optical table (a layout of the experimental setup is available in the supplementary material).

A bidisperse granular bed of height 24 mm $\leq$ $h$ $\leq$ 25 mm was set up in the flume, and the remaining space was filled with a viscous liquid whose refractive index was matched with that of the bed. The bed consisted of larger and smaller glass spheres (soda–lime–silica glass) with density $\rho_s$ = 2500 kg/m$^3$ and diameters $d_1$ = 3 mm $\pm$ 0.2 mm and $d_2$ = 2 mm $\pm$ 0.2 mm, respectively, and which we call species 1 and 2. The ratio of the total volume occupied by the small spheres $V_2$ to that of large ones $V_1$ was $V_2/V_1$ = 1.5, and we considered the mean diameter as $d$ $=$ $(0.4d_1 + 0.6d_2) /2$ $=$ 2.4 mm (i.e., averaged by the mass proportion of each species). For the fluid, we used an oil for fluorescence microscopy from Cargille Laboratories, with dynamic viscosity $\mu$ = 651 cP, density $\rho$ = 931 kg/m$^3$, and refractive index (for 532 nm) of 1.5127 at 23 $^\circ$C, which assured the desired shear while matching the refractive index of grains. The fluid viscosity was measured with a rheometer Anton Paar MCR 102, showing that it was within 770 and 800 cSt during the experiments, and the room temperature was 22$^o$C $\pm$ 1$^o$C during all tests. Tables showing the composition used in each test and microscopy images of the used grains are available in the supplementary material.

With that, a liquid film above the granular bed, with height 5.7 mm $\leq$ $h_f$ $\leq$ 6.8 mm, was sheared by the rotating lid at a constant angular velocity $\Omega$, creating a laminar Couette flow. The angular velocities $\Omega$ varied within 3 and 6 rpm, corresponding to lid velocities at its centerline of 53 mm/s $\leq$ $U_{lid}$ $\leq$ 106 mm/s, shear rates of 8.1 s$^{-1}$ $\leq$ $\dot{\gamma}$ = $U_{lid}/h_f$ $\leq$ 19.5 s$^{-1}$, and Shields numbers of 0.14 $\leq$ $\theta$ $\leq$ 0.35, where $\theta$ = $\mu \dot{\gamma} ((\rho_s - \rho )g d)^{-1}$, $g$ = $|\vec{g}|$ being the modulus of the acceleration of gravity. The critical value of the Shields number, $\theta_{c}$, was fixed at 0.1 (estimated from values found in the literature \cite{Houssais_1, Houssais_2, Ferdowsi}). The Reynolds number based on the fluid height, Re = $\rho U_{lid} h_f \mu^{-1}$, was within 0.46 and 0.92, and the Reynolds number based on the mean grain diameter, Re$_{p}$ = $\rho \dot{\gamma} d^2 \mu^{-1}$, within 0.05 and 0.23. The ratio between the grain and fluid densities was $S$ = $\rho_s / \rho$ = 2.7. Prior to each test, the upper lid was rotated at $\Omega$ = 25 rpm ($\theta$ = 1.5) during 60 seconds in order to suspend all the grains with the exception of the bottom-most layers, followed by a rest period of 5 minutes for the grains to settle.

One continuous 0.2 W laser head emitting at 532 nm was mounted over and another one below the flume, both generating a vertical plane traversing the bed and forming a single laser sheet (approximately 1 mm thick). We used two lasers for having a regular distribution of light through the bed (the bed being lighted from both its top and bottom). A digital camera with a lens of 18--140 mm focal distance and F2.8 maximum aperture was mounted with a perpendicular view to the laser sheet. The camera was of complementary metal-oxide-semiconductor (CMOS) type with a maximum resolution of 20.9 Mpx for photographs and 1920 px $\times$ 1080 px at 60 Hz for movies. The regions of interest (ROIs) were set at 1920 px $\times$ 940 px for the movies and 2780 px $\times$ 1410 px for photographs, for a field of view of 60 mm $\times$ 30 mm. Movies were recorded at 30 Hz for the first 40 or 80 min of experiments in order to capture segregation within the bedload layer during the very first stages of the flow, and images were acquired at 0.05 Hz during 140 hours in order to sample the slow segregation and compaction in the solid-like region. In addition, movies were recorded at 30 Hz for 10 min every 4 hours in order to accurately measure the changes that occur between the bedload and solid-like layers. Afterward, the movies and images were processed by a code written in the course of this work. Movies showing the evolution of the bed are available in the supplementary material, and the image-processing codes and images are available in an open repository \cite{Supplemental}. More details about particle detection and computation of velocities, packing fraction and strain are available in the supplementary material.

\section{RESULTS}
\label{sec:results}

\begin{figure}[h!]
	\begin{center}
		\includegraphics[width=.65\linewidth]{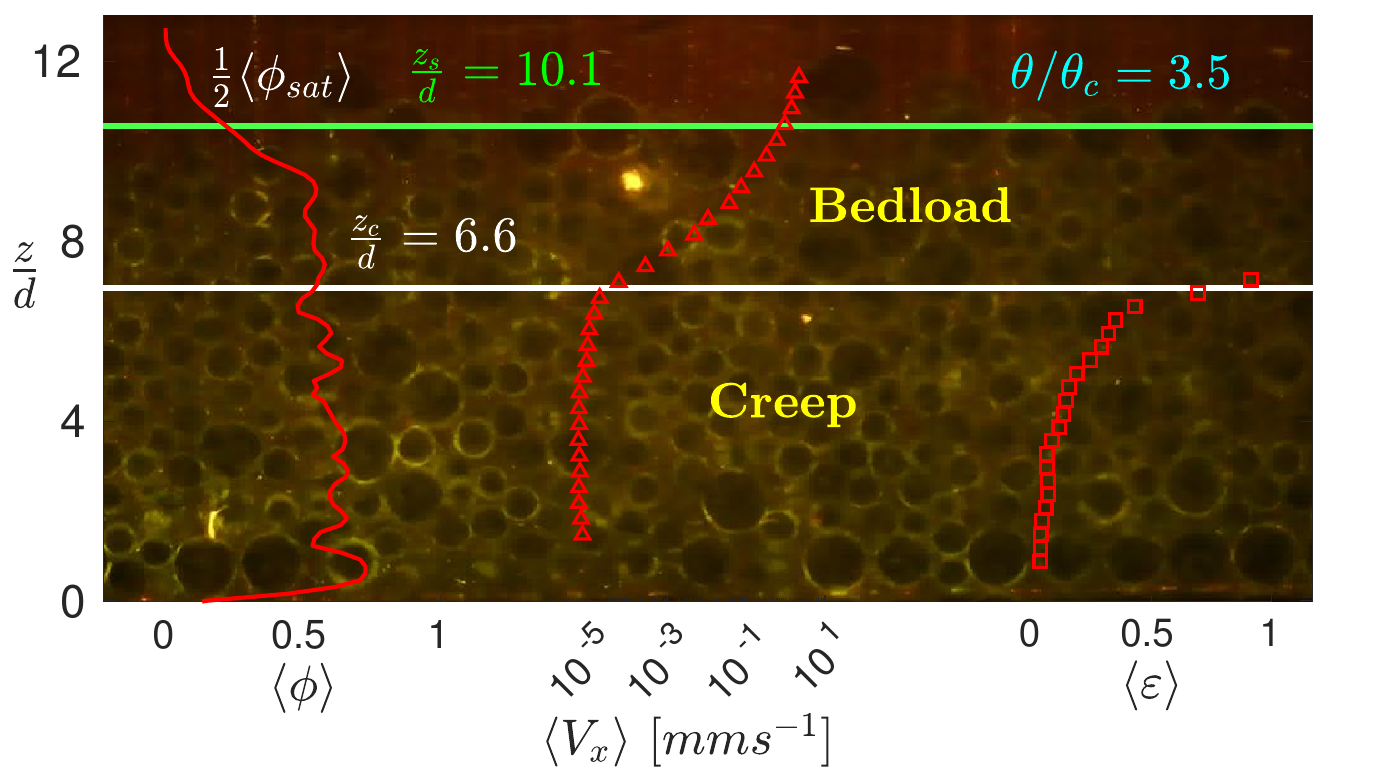}\\
	\end{center}
	\caption{Spatio-temporal averages of packing fraction $\left< \phi \right>$, velocity $\left< V \right>$ and strain $\left< \varepsilon \right>$ within the bed for $\theta / \theta_c$ = 3.5. In the image, flow is from left to right.}
	\label{fig:bed_general}
\end{figure}

As soon as the lid begins moving, the fluid entrains grains into motion, with grains near the surface moving as bedload while those below move as creep (with velocities much smaller than those of bedload grains). With the fluid and velocities used, topmost grains of the bedload layer moved by rolling and sliding over other grains, and there was no grain in suspension. By processing the acquired images and movies, we computed spatio-temporal averages of the packing fraction $\left< \phi \right>$, longitudinal velocity $\left< V \right>$ and strain $\left< \varepsilon \right>$ within the bed. The time averages were computed within specific intervals (as indicated in the following) and space averages were computed only in the longitudinal direction (not in the vertical, unless otherwise specified), so that we ended with vertical profiles of $\left< \phi \right>$, $\left< V \right>$ and $\left< \varepsilon \right>$ for different applied stresses. For example, Fig. \ref{fig:bed_general} shows $\left< \phi \right>$, $\left< V \right>$ and $\left< \varepsilon \right>$ (with the bed as background) for $\theta / \theta_c$ = 3.5. We note the existence of oscillations with wavelength of the order of $d$ in the vertical profiles of $\left< \phi \right>$, which are due to the settling of particles in layers \cite{Houssais_2}. Details of the computations of $\left< \phi \right>$, $\left< V \right>$ and $\left< \varepsilon \right>$ are available in the supplementary material.

Based on the profiles of $\left< \phi \right>$ and $\left< V \right>$, we determined the regions where creeping and bedload take place by computing the vertical positions $z_c$ and $z_s$. The position $z_s$ is defined as that where $\left< \phi \right>$ = $1/2 \left< \phi_{sat} \right>$, so that for $z > z_s$ the concentration of grains is low and bedload vanishes, while $z_c$ is the position where a kink takes place in the $\left< V \right>$ profiles, separating the regions where creep ($z < z_c$) and bedload ($z_c < z < z_s$) occur \cite{Houssais_1, Cunez2}. In our experiments, this kink (and thus $z_c$) always corresponded to the position where the viscous number $I_v$ is approximately 2 $\times$ 10$^{-8}$.  The viscous number $I_v$ is the ratio between the microscopic and macroscopic timescales \cite{GDR_midi} applied to viscous flows,

\begin{equation}
	I_v = \frac{\mu \dot{\gamma}}{P_p} \,\,,
	\label{eq:Iv}
\end{equation}

\noindent where $P_p$ is the confinement pressure. This pressure decreases with height, being the result of the load of material above the considered height $z$. We computed $P_p$ as in Houssais et al. \cite{Houssais_2},

\begin{equation}
	P_p = \left( \rho_s - \rho \right) g \left[ \frac{\forall_s}{A_{cont}} +  \int_{z}^{\infty} \left< \phi \right> dz \right] \,\,,
	\label{eq:Pp}
\end{equation}	

\noindent where $\forall_s$ is the volume of one grain (computed using the mean diameter $d$) and $A_{cont}$ is the characteristic surface of contact between a typical topmost grain and the bed surface. We proceeded as in Houssais et al. \cite{Houssais_2} and considered that $\forall_s / A_{cont}$ is equal to an integration constant $\alpha$ = 0.1$d$. Finally, the effective viscosity $\mu_{eff}$ is computed by Eq. \ref{eq:mu_eff},

\begin{equation}
	\mu_{eff} = \frac{\tau}{\dot{\gamma}} \,\,,
	\label{eq:mu_eff}
\end{equation}

\noindent where $\tau$ is the applied stress.

\begin{figure}[h!]
	\begin{center}
		\includegraphics[width=.8\linewidth]{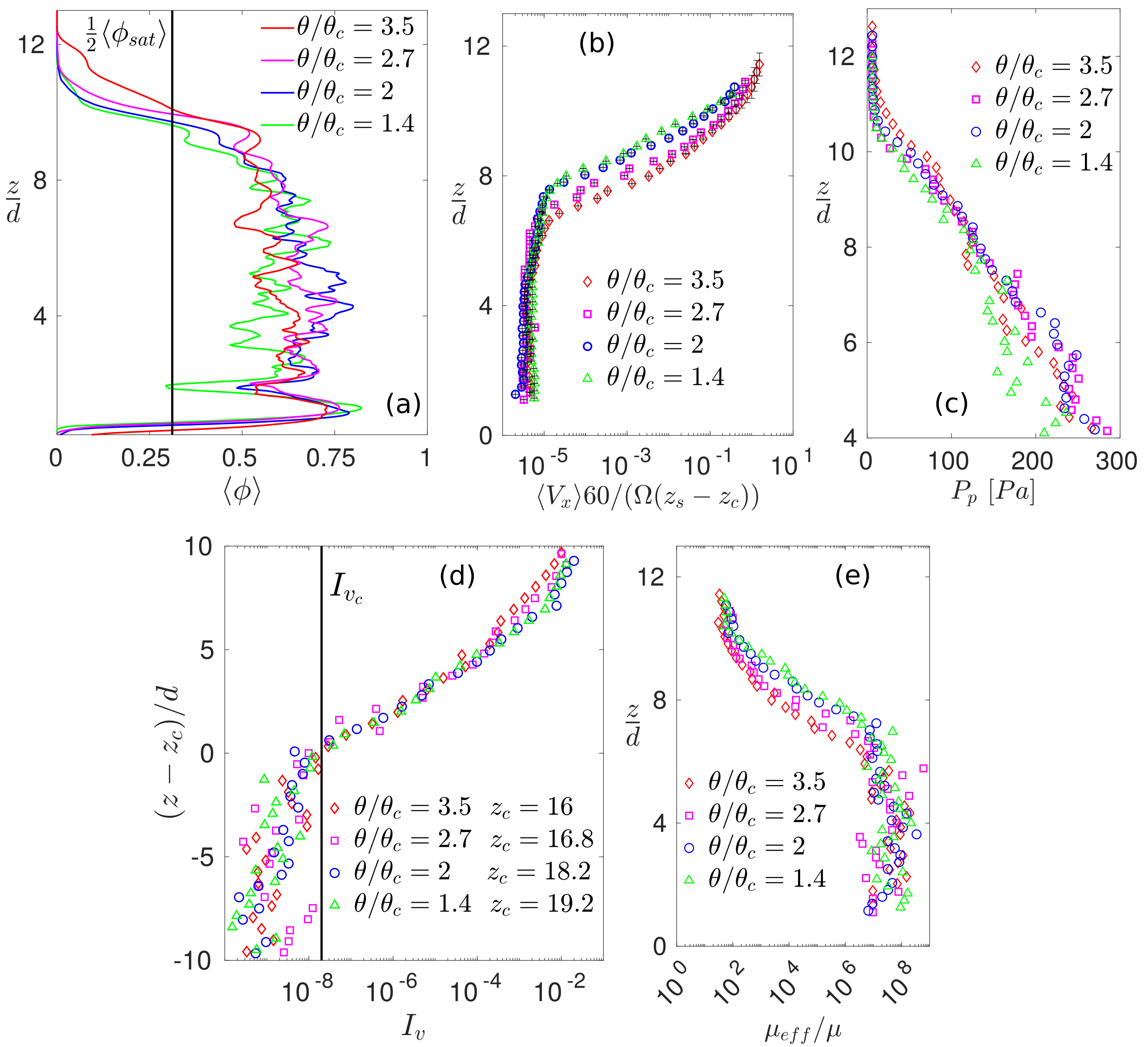}\\
	\end{center}
	\caption{Vertical profiles of averaged (a) packing fraction $\left< \phi \right>$, (b) grain velocity $\left< V \right>$ normalized by that of the lid, (c)  confinement pressure $P_p$, (d) viscous number $I_v$, and (e) effective viscosity $\mu_{eff}$ normalized by that of the fluid, for different Shields numbers $\theta$. In panel (d), the vertical coordinate has its origin in the creep-bedload transition.}
	\label{fig:hardening2}
\end{figure}

Figures \ref{fig:hardening2}a--\ref{fig:hardening2}e show vertical profiles of space-time averaged packing fraction $\left< \phi \right>$, grain velocity $\left< V \right>$, confinement pressure $P_p$, viscous number $I_v$, and effective viscosity $\mu_{eff}$, for different Shields numbers $\theta$. The profiles of $\left< V \right>$ show that velocities are higher on the bed surface ($z/d$ = 10-12), decrease relatively fast with depth in the bedload layer (7-8 $\leq$ $z/d$ $<$ 10-12), and have much lower values (5 orders of magnitude lower than on the bed surface) and a smoother decrease with depth in the creep layer ($z/d$ $<$ 7-8), a kink existing in the transition from creep to bedload. This kink occurs at a height $z$ where $I_v$ $\approx$ 2 $\times$ 10$^{-8}$, as can be seen in Fig. \ref{fig:hardening2}d. This figure shows that $I_v$ is maximum at the bed surface, decreases strongly with depth in the bedload layer and smoothly in the creep layer, with the kink occurring at $z$ = $z_c$. For the packing fraction $\left< \phi \right>$, we observe a fast increase with depth in the bedload layer, with an average constant value in the creeping layer. As noted for Fig. \ref{fig:bed_general}, oscillations with a wavelength of the order of $d$ are present, which are due to the settling of particles in layers. On the bottom, $\left< \phi \right>$ tends to zero since the contact area between the spherical particles and the channel wall is very small. The effective viscosity $\mu_{eff}$ increases with depth, from values of the order of 10$\mu$ at the bed surface to 10$^7 \mu$ at $z$ = 6-7, from which depth it remains constant until reaching the bottom. This indicates a solid-like behavior in the creeping layer. Finally, the pressure $P_p$ is roughly constant for 10 $\leq$ $z/d$ $<$ 12 (top region of the bedload layer), and increases with depth for $z/d$ $<$ 10. Those results are in agreement with the experiments of Houssais et al. \cite{Houssais_1, Houssais_2}, carried out with much lighter grains ($S$ = 1.1) than our experiments ($S$ = 2.7). For that reason, the magnitude of pressures at the bottom of the bed are two order of magnitude higher in our experiments when compared with those in Houssais et al. \cite{Houssais_1, Houssais_2}.

From the ensemble of experiments, we observed particle segregation and strain hardening, which we discuss next.

\subsection{Segregation}
\label{sec:segregation}

\begin{figure}[h!]
	\begin{center}
		\includegraphics[width=.99\linewidth]{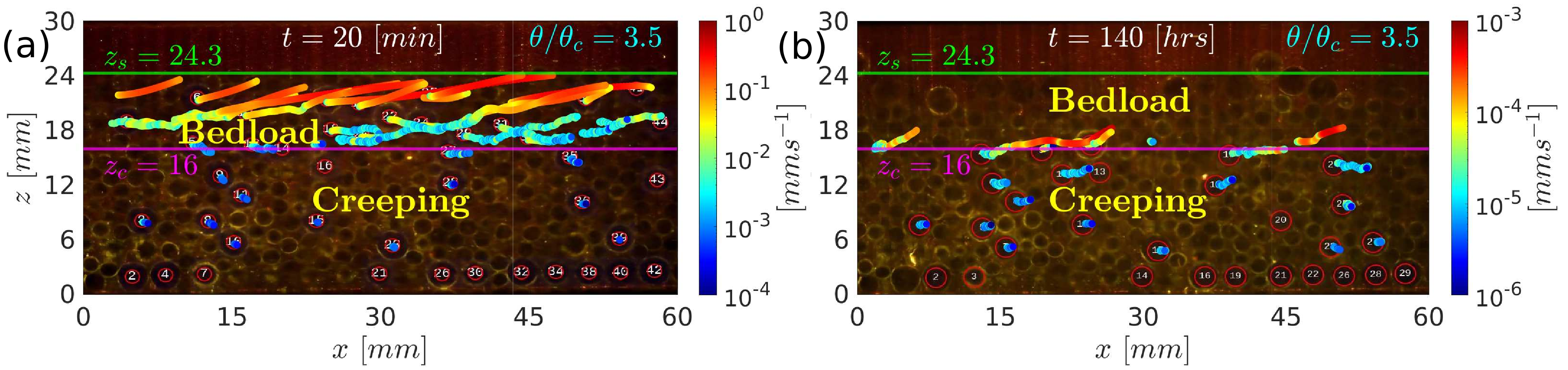}\\
	\end{center}
	\caption{Motion of the large particles in the segregation zone. (a) Migration of large particles from $t$ = 0 to 20 min; (b) migration of large particles from $t$ = 20 min to 140 h. The graphics correspond to $\theta / \theta_c$ = 3.5. Multimedia available online.}
	\label{fig:segregation1}
\end{figure}

We begin with the segregation, for which we followed the motion of the large particles appearing in the recorded images (with an accuracy of approximately 0.005 mm, by using subpixel methods). We assigned a label to each one of those particles and tracked them along the movie frames and photographs. For example, Fig. \ref{fig:segregation1}a shows the trajectories of large particles that were segregating during the first 20 min of tests and Fig. \ref{fig:segregation1}b for $t$ = 20 min to 140 h, both for $\theta / \theta_c$ = 3.5 (multimedia available online). We observe that displacements during segregation are much higher in the bedload layer when compared with the creep layer (corroborating our computations of $z_c$). We also notice that most of segregation, given by the vertical motion of large particles, occurs within the bedload layer, and that the intensity of segregation is much higher during the first 20 minutes than during the next 139h 40 min.

\begin{figure}[h!]
	\begin{center}
		\includegraphics[width=.6\linewidth]{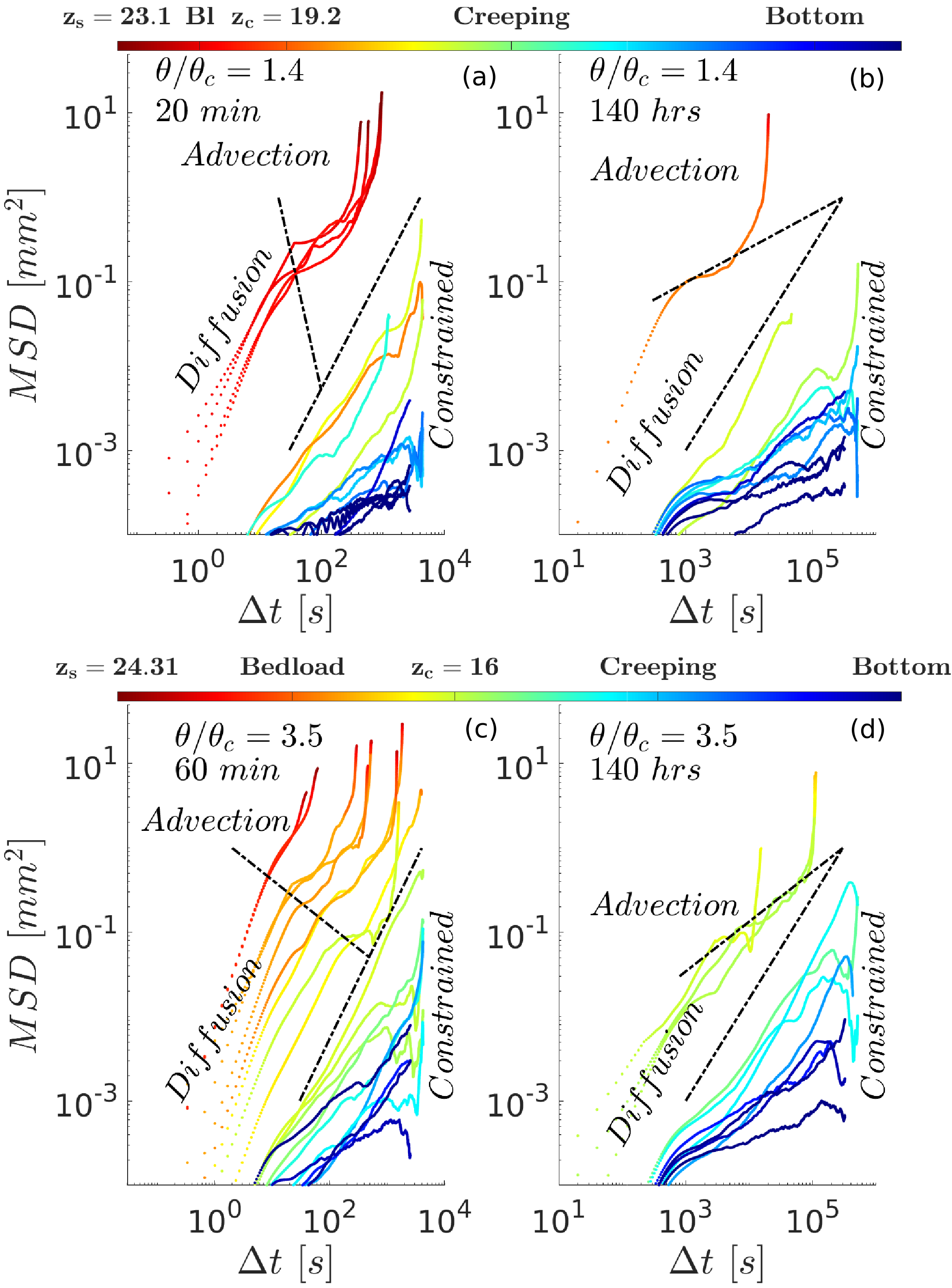}\\
	\end{center}
	\caption{Mean Squared Displacement (MSD) of large particles for (a) $\theta / \theta_c$ = 1.4 and from $t$ = 0 to 20 min; (b) $\theta / \theta_c$ = 1.4 and from $t$ = 20 min to 140 h; (c) $\theta / \theta_c$ = 3.5 and from $t$ = 0 to 20 min; (d) $\theta / \theta_c$ = 3.5 and from $t$ = 20 min to 140 h. The times indicated in each panel correspond to the duration used to compute the MSDs, and \textit{Bl} in panels (a) and (b) stands for \textit{bedload}. Dots are colored according to the instantaneous positions of particles (read in the colorbar).}
	\label{fig:segregation2}
\end{figure}

In order to quantify the intensity of displacements in the vertical direction and the regions where large grains move, we computed the Mean Squared Displacement \cite{Michalet, Ernst} (MSD) of large particles,

\begin{equation}
	MSD(\Delta  t) = \frac{1}{N} \sum^{N} \left[ z(t + \Delta  t) - z(t) \right] ^2 \,\,,
	\label{eq:MSD}
\end{equation}

\noindent where $N$ is the number of averaged points, $\Delta  t$ is the interval for a given MSD computation, and $MSD(\Delta  t)$ corresponds to the area visited by the considered particle during the interval $\Delta  t$. In addition to mean distances traveled by the considered particle, the curves of $MSD$ as functions of $\Delta  t$ inform about regions where the particle is advected, moves by pure diffusion, or is confined. Typically, this kind of plot is curved upwards in case of advection (superdiffusion), is a straight line in case of pure diffusion, and is curved downwards in case of confinement (subdiffusion) \cite{Saxton, Ferdowsi}. We note that MSD is used here as an analogy for hints about the behavior of our system, MSD being typically used for more homogeneous systems with smaller particles. MSD was used with granular beds in previous works for the same purpose \cite{Ferdowsi}, but care must be taken.

Figure \ref{fig:segregation2} shows the $MSD$ as a function of $\Delta  t$ for the large particles, Fig. \ref{fig:segregation2}a corresponding to $\theta / \theta_c$ = 1.4 from $t$ = 0 to 20 min, Fig. \ref{fig:segregation2}b to $\theta / \theta_c$ = 1.4 from $t$ = 20 min to 140 h, Fig. \ref{fig:segregation2}c to $\theta / \theta_c$ = 3.5 from $t$ = 0 to 20 min, and Fig. \ref{fig:segregation2}d to $\theta / \theta_c$ = 3.5 from $t$ = 20 min to 140 h. These graphics show that the advection of large particles occurs exclusively in the bedload layer and are more intense during the first 20 minutes, while pure diffusion is seen to occur in the creep layer close to the limit with the bedload layer, being more intense during the first 20 min, but also occurring considerably at later times. In the lower region of the creep layer (farther from the bedload boundary), large particles are seen to be confined.

\begin{figure}[h!]
	\begin{center}
		\includegraphics[width=.65\linewidth]{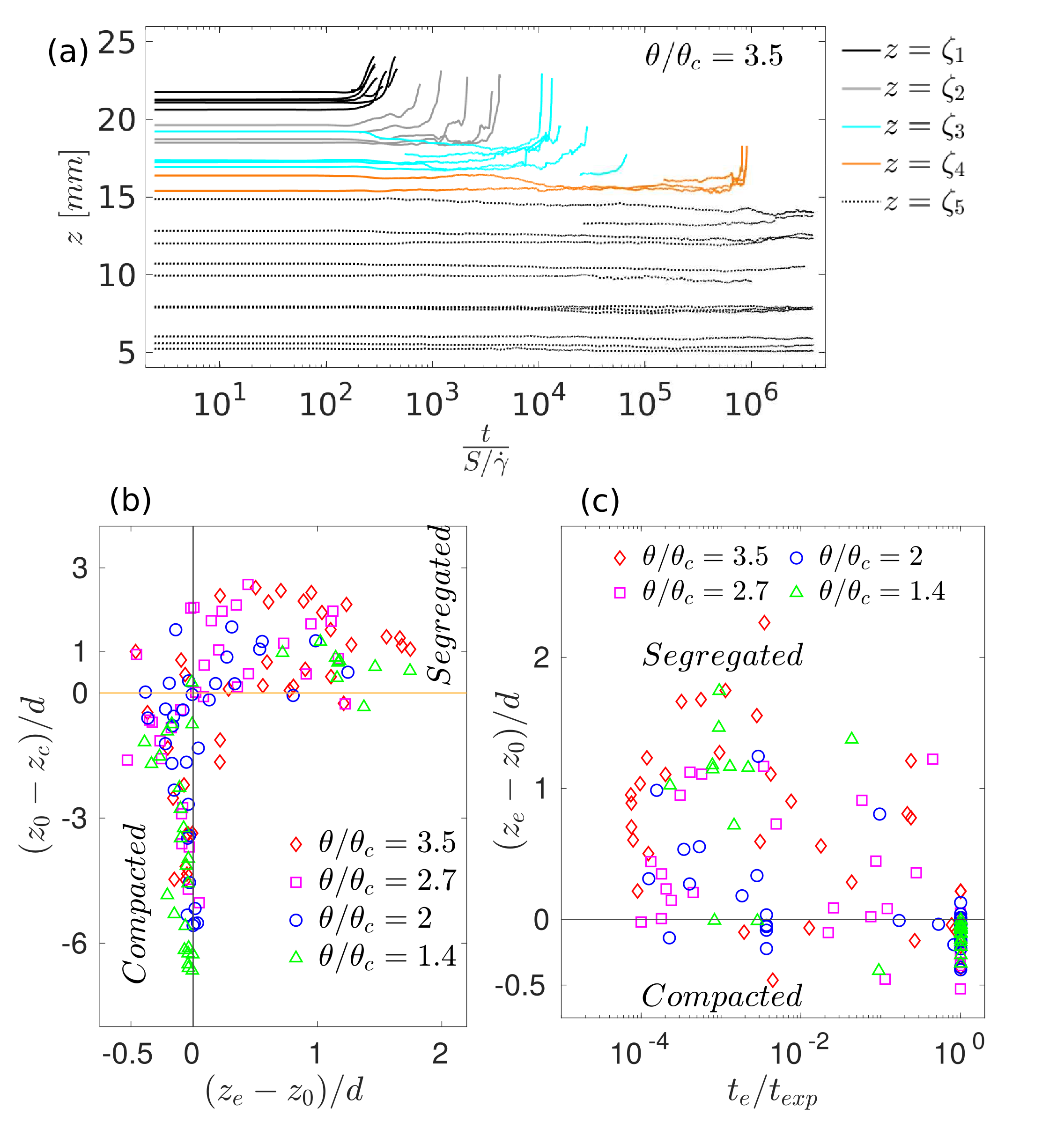}\\
	\end{center}
	\caption{(a) Vertical position $z$ of the large particles that segregated (or were deeper in the bed) along time, for $\theta / \theta_c$ = 3.5. In the figure, time is normalized by $t_{shear}$ = $S/\dot{\gamma}$, and different colors and line types are used for curves showing particles originally in different regions, where: 0.85$z_s$ $\leq$ $\zeta_1$ $\leq$ $z_s$; 0.75$z_s$ $\leq$ $\zeta_2$ $<$ 0.85$z_s$; $z_c$ $\leq$ $\zeta_3$ $<$ 0.75$z_s$; and 0.95$z_c$ $\leq$ $\zeta_4$ $<$ $z_c$. A dimensional form of the diagram is available in the supplementary material. (b) Initial position $z_0$ of large particles (with origin at $z_c$) as a function of their vertical displacement $z_e - z_0$, for different shear stresses. In the graphic, $z_e$ is the final position of particles and both axes are normalized by $d$.  (c) Vertical displacement $z_e - z_0$ of large particles as a function of the instants when they were last detected $t_e$, for different shear stresses. The instants $t_e$ are normalized by the total time of experiments $t_{exp}$. Panels (b) and (c) indicate the regions where segregation has effectively occurred and those in which compaction has taken place.}
	\label{fig:segregation3}
\end{figure}

Figure \ref{fig:segregation3}a shows the vertical position $z$ of the large particles that segregated (or were deeper in the bed) along time, for $\theta / \theta_c$ = 3.5 (graphics in dimensional form and for the other shear stresses are available in the supplementary material). In Fig. \ref{fig:segregation3}a, the time is normalized by $t_{shear}$ = $S/\dot{\gamma}$. We observe that the segregation times (in logarithmic scale in the graphic) differ according to the vertical position the particle is initially at. Therefore, Fig. \ref{fig:segregation3}a shows in different colors and line types the curves corresponding to particles originally in different regions: 0.85$z_s$ $\leq$ $\zeta_1$ $\leq$ $z_s$; 0.75$z_s$ $\leq$ $\zeta_2$ $<$ 0.85$z_s$; $z_c$ $\leq$ $\zeta_3$ $<$ 0.75$z_s$; and 0.95$z_c$ $\leq$ $\zeta_4$ $<$ $z_c$. In terms of order of magnitude, the large particles in the $\zeta_1$ region segregate until $t/t_{shear}$ $\sim$ 10$^2$ (within the first minute), those in the $\zeta_2$ region within 10$^2$ and 10$^3$ (within 1 and 10 minutes), and those in the $\zeta_3$ region within 10$^3$ and 10$^4$ (within 10 and 100 minutes), these three regions corresponding to the bedload layer. Particles originally in the $\zeta_4$ region, which corresponds to the upper part of the creep layer (in the vicinity of the bedload layer), segregate within $t/t_{shear}$ $\sim$ 10$^5$ and 10$^6$ (within 100 and 1000 minutes). Below this region, we have not observed segregation within the duration of our experiments.

Figure \ref{fig:segregation3}b shows the initial position $z_0$ of the large particles of Fig. \ref{fig:segregation3}a (with origin at $z_c$) normalized by $d$ as a function of their vertical displacement $\left( z_e - z_0 \right)/d$, for different shear stresses, where $z_e$ is the final position of particles. During the experiments, we have not remarked large particles moving distances of the order of their radius in the transverse direction, so that they did not leave completely the laser plane (i.e., information for identifying $z_e$ was complete). Figure \ref{fig:segregation3}b also indicates the regions where segregation has effectively occurred (upwards motion of large particles) and those in which compaction has taken place (collective downward motion of all particles). Since the origin of the final position (ordinate) is $z_c$, we notice that most of compaction takes place in the creep layer, while most of segregation occurs in the bedload layer. In addition, we observe that segregation is stronger for higher shear stresses while compaction is stronger for lower shear stresses. Figure \ref{fig:segregation3}c shows the vertical displacements of the large particles of Fig. \ref{fig:segregation3}a $\left( z_e - z_0 \right)/d$ as a function of the instants when they were last detected $t_e/t_{exp}$, where $t_{exp}$ is the duration of each experiment. We note that the data concentration at $t_e/t_{exp}$ = 1 is due to particles that were detected until the end of experiments.

\begin{figure}[h!]
	\begin{center}
		\includegraphics[width=.9\linewidth]{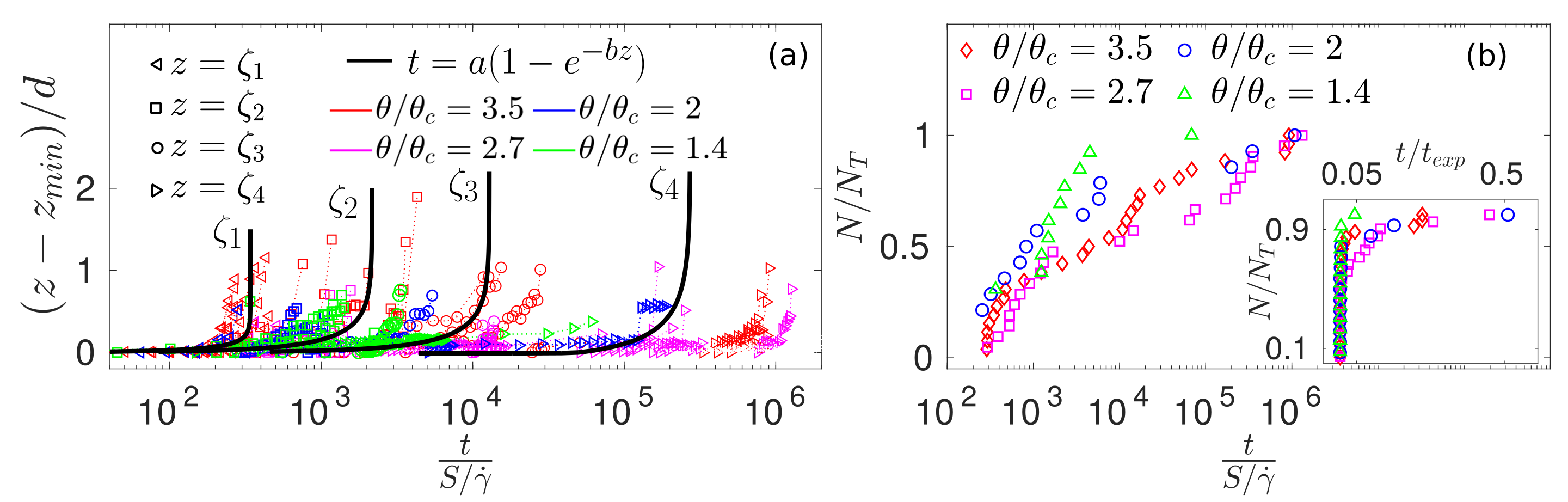}\\
	\end{center}
	\caption{(a) Displaced position $\left( z_ - z_{min} \right)/d$ of the large particles that segregated (symbols) and fittings (black lines) of the corresponding averages as functions of $t/t_{shear}$, for each region where segregation takes place and different shear stresses. Fittings follow exponential functions (as proposed by Zhou et al. \cite{Zhou2} for the degree of segregation), and $z_{min}$ is the lowest position reached by the particle (due to an increase in bed compaction) before start rising. (b) Number of segregated particles $N$ normalized by the total number of large particles $N_T$ as a function of $t/t_{shear}$, for different shear stresses. The inset shows $N/N_t$ as a function of $t/t_{exp}$. Dimensional forms of panels (a) and (b), and graphics of $dN/dt$ as a function of $t$ are available in the supplementary material.}
	\label{fig:segregation4}
\end{figure}

Finally, Fig. \ref{fig:segregation4}a shows the displaced position of the large particles that segregated (symbols), $\left( z_ - z_{min} \right)/d$, and fittings (black lines) of the corresponding averages as functions of $t/t_{shear}$, for each region where segregation takes place and different shear stresses. In this displaced coordinate, $z_{min}$ is the lowest position reached by the particle (due to an increase in bed compaction) before start rising, and the fittings follow exponential functions (as proposed by Zhou et al. \cite{Zhou2} for the degree of segregation). Figure \ref{fig:segregation4}b shows the number of segregated particles $N$ normalized by the total number of large particles $N_T$ identified in the images as a function of $t/t_{shear}$, for different shear stresses (segregation rates can be obtained by taking the time derivative of those curves, $dN/dt$, and are available in the supplementary material). From Fig. \ref{fig:segregation4}a, we observe a consistent behavior for the shear stresses tested, with similar segregation curves for each depth ($\zeta_1$ to $\zeta_4$) and with no clear dependency on $\theta$, although varying with it. The timescales to complete the segregation in each region are the same observed above for Fig. \ref{fig:segregation3}a. Figure \ref{fig:segregation4}b shows that the number of segregated particles vary with $\theta$, with periods of high slope alternating with others of low slope in the graphic. Although no clear tendency with $\theta$ can be found, the general behavior of curves is similar. The curves have initially a high slope, then a significant decrease in the slope occurs in a time that depends on the shear stress, the slope increases again to approximately the previous values after some time has elapsed, and, finally, by the end of the experiments, the slope decreases again. We do not have an explanation for these oscillations, but they could vary with the characteristic time of segregation of each region. For example, for $\theta / \theta_c$ = 3.5 a high slope is observed in Fig.  \ref{fig:segregation4}b for $t/t_{shear}$ $\sim$ 10$^2$ ($t$ $\sim$ 1 min), then a small slope for $t/t_{shear}$ $\sim$ 10$^3$ ($t$ $\sim$ 10 min), a high slope again for $t/t_{shear}$ $\sim$ 10$^4$ ($t$ $\sim$ 10$^2$ min), and a low slope for $t/t_{shear}$ $\sim$ 10$^5$--10$^6$ ($t$ $\sim$ 10$^3$ min). This remains, however, to be investigated further. Lastly, the inset in Fig. \ref{fig:segregation4}b shows $N/N_t$ as a function of $t_{seg}/t_{exp}$, from which we observe that a great part of segregation occurs in the very beginning of experiments (within the first 1\% of the total time), so that before 0.5 $t_{exp}$ more than 90\% of larger grains have already segregated.

To summarize, we found a characteristic time for the upward motion (segregation) of large particles (Figs. \ref{fig:segregation3}a and \ref{fig:segregation4}a) which depends on the depth within the bed:

\begin{itemize}
	\item $t_1/t_{shear}$ = 10$^2$ ($t_1$ = 1 min), for 0.85$z_s$ $\leq$ $\zeta_1$ $\leq$ $z_s$;
	\item $t_2/t_{shear}$ = 10$^3$ ($t_2$ = 10 min), for 0.75$z_s$ $\leq$ $\zeta_2$ $<$ 0.85$z_s$;
	\item $t_3/t_{shear}$ = 10$^4$ ($t_3$ = 10$^2$ min), for $z_c$ $\leq$ $\zeta_3$ $<$ 0.75$z_s$;
	\item $t_4/t_{shear}$ = 10$^5$--10$^6$ ($t_3$ = 10$^3$ min), for 0.95$z_c$ $\leq$ $\zeta_4$ $<$ $z_c$.
\end{itemize}

\noindent The first three regions ($\zeta_1$ to $\zeta_3$) correspond to the bedload layer, where large particles move mainly by advection (Fig. \ref{fig:segregation2}), while the the last one ($\zeta_4$) corresponds to the upmost part of the creep layer (within 95\% of its top), where particles move mainly by diffusion (Fig. \ref{fig:segregation2}). In this top layer, creep seems strongly influenced by the shear caused by the above bedload layer. Below in the creep layer ($z$ $<$ 0.95$z_c$), we have not observed any upward motion of large particles, but, instead, a collective downward motion due to bed compaction (Fig. \ref{fig:segregation3}). This increase in bed compaction results in strain hardening and decrease in the the granular mobility \cite{Cunez2}, which are investigated next (in Subsection \ref{sec:hardening}). In this picture, large particles in the upmost part of the creep layer move slowly by diffusion until reaching the creep-bedload transition zone ($z$ = $z_c$), from which height they are vertically advected by the rapid motion of surrounding particles, the vertical velocity increasing with height.

\subsection{Strain hardening}
\label{sec:hardening}

\begin{figure}[h!]
	\begin{center}
		\includegraphics[width=.9\linewidth]{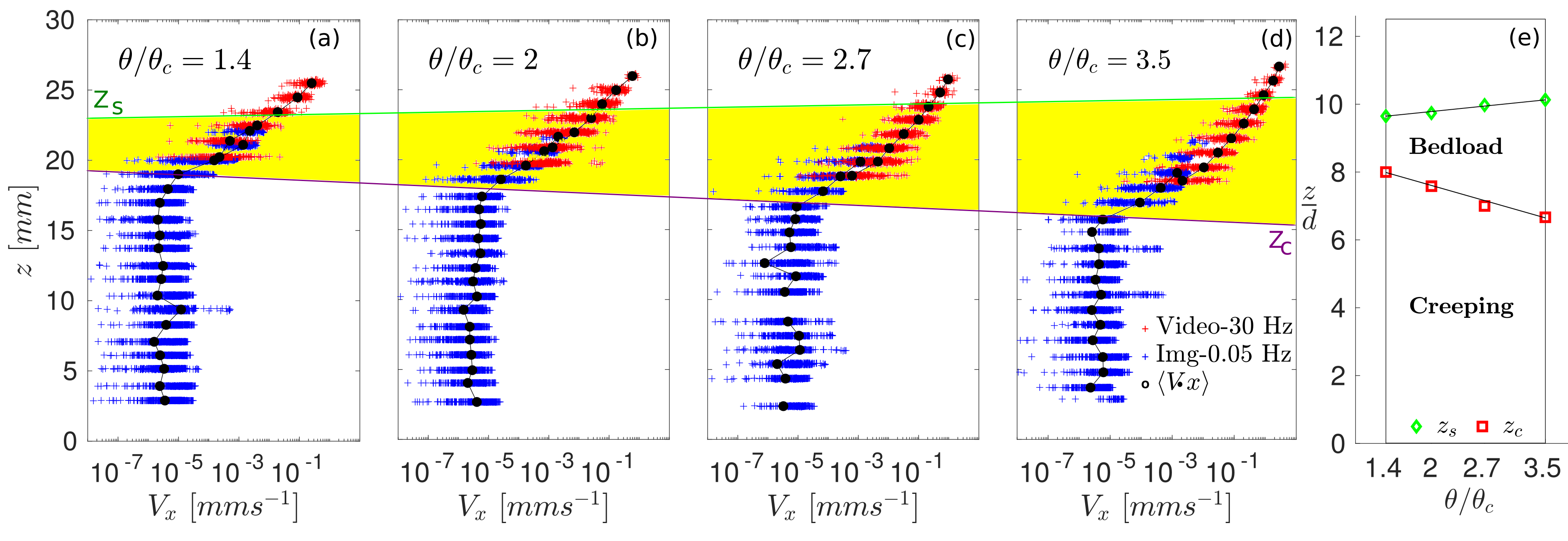}\\
	\end{center}
	\caption{(a)-(d): Instantaneous longitudinal velocity at each height $V_x(z)$, for all detected grains and different shear stresses $\theta / \theta_c$. Red symbols were computed from movies and blue symbols from photographs, black dots are simple averages for a given height, and the yellow region corresponds to the bedload layer. (e) Heights $z_c$ and $z_s$ as functions of $\theta / \theta_c$. Multimedia available online.}
	\label{fig:hardening1}
\end{figure}

Figures \ref{fig:hardening1}a-d (multimedia available online) show the longitudinal component of the instantaneous velocity measured for all detected particles and the entire duration of each test as a function of the bed height $z$, for different shear stresses (Figs. \ref{fig:hardening1}a-\ref{fig:hardening1}d correspond to $\theta / \theta_c$ from 1.4 to 3.5, respectively). We observe that, even considering the entire duration of tests, the bed behavior is consistent, with a region where velocity gradients are higher and which corresponds to the bedload layer, and another one where gradients are much lower and which corresponds to the creep layer. In particular, we observe that by increasing the shear stress, the bedload layer increases, which is the layer where bed dilation occurs, in agreement with previous works \cite{Cunez2}. This is also the layer where the vertical advection of large particles takes place, so that segregation is stronger. The magnitude of longitudinal velocities also increases with the shear stress, going from roughly 0.25 mm/s for the topmost grains when $\theta / \theta_c$ = 1.4 to 2.5 mm/s when $\theta / \theta_c$ = 3.5, contributing then to higher segregation rates. On the contrary, the creep layer shortens with increasing the shear stress, with average velocities of the order of 10$^{-6}$--10$^{-5}$ mm/s. This is the layer where compaction takes place over the time, and both isotropic and anisotropic hardenings occur \cite{Cunez2}. Figure \ref{fig:hardening1}e presents the heights $z_c$ and $z_s$ as functions of $\theta / \theta_c$, showing that the bedload and creep layers increase and decrease, respectively, linearly with the shear stress. At the leading order, this reflects a quadratic variation of the bedload flow rate with the applied shear stress if both the particles' velocities and bedload height vary linearly with $\theta$. This is, indeed, a reasonable picture for laminar viscous flows in which both Re and Re$_p$ $<$ 1, Charru et al. \cite{Charru_6} having shown that $V_x$ $\sim$ $\theta$ and that the bedload flow rate varies with $\theta ^2$ when $\theta$ is close to $\theta_c$ (explaining then the linear variation of the bedload layer). However, nonlinearities due to bidispersity and deviations from the critical conditions are expected.

\begin{figure}[h!]
	\begin{center}
		\includegraphics[width=.75\linewidth]{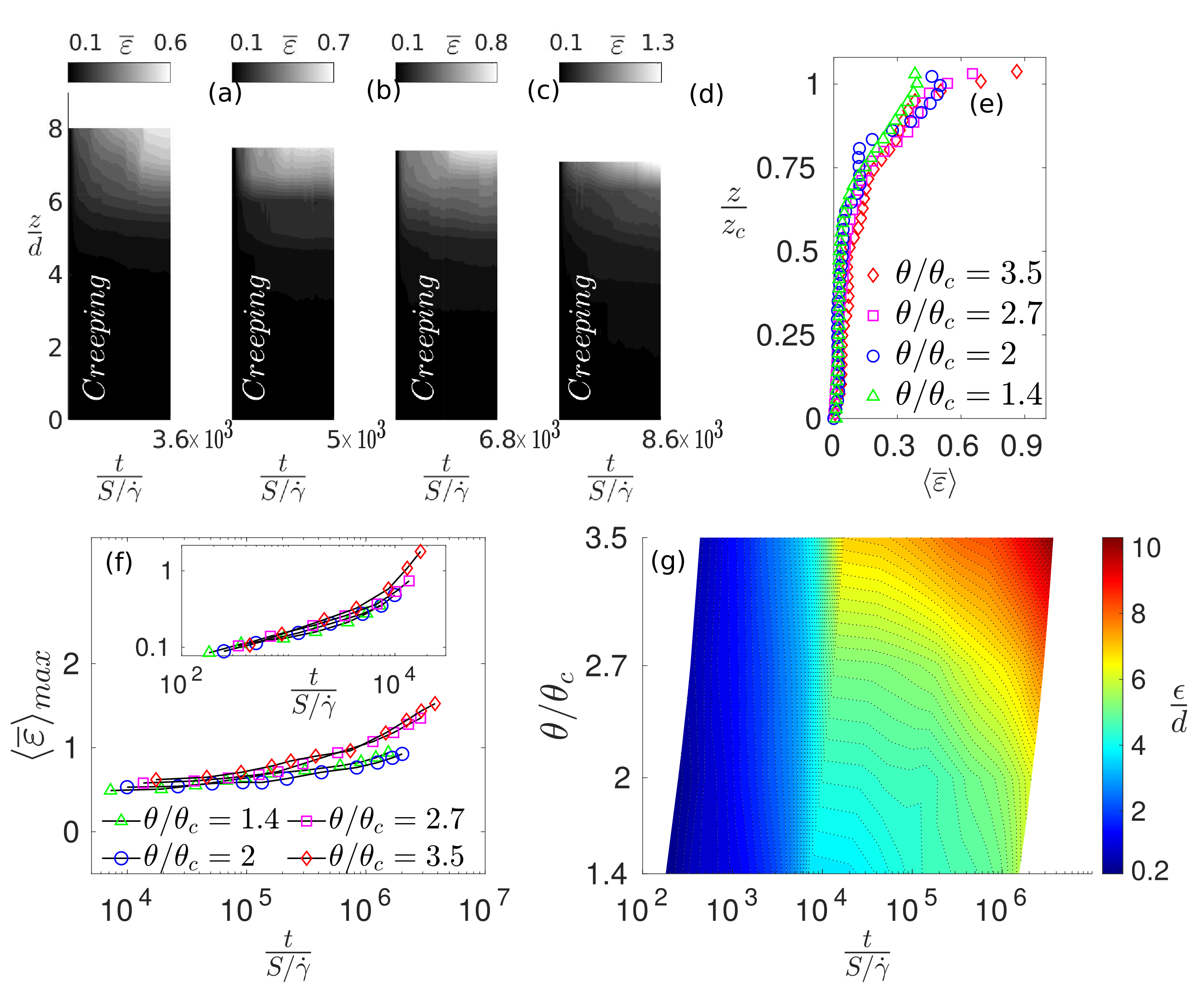}\\
	\end{center}
	\caption{(a)--(d) Space-time diagrams of the longitudinal average strain $\overline{\varepsilon}$ for, respectively, $\theta / \theta_c$ from 1.4 to 3.5 (from left to right). (e) Vertical profiles of the time-longitudinal average of the strain $\left< \overline{\varepsilon} \right>$ within the creep layer for different shear stresses. (f) Time evolution of the maxima of time-longitudinal averages of the strain, $\left< \overline{\varepsilon}\right>_{max}$, for different shear stresses $\theta / \theta_c$. The inset corresponds to the first 40 min. (g) Map in the $\theta / \theta_c$ vs. $t/t_{shear}$ space of the integrals of $\left< \overline{\varepsilon} \right>$ profiles, $\epsilon$, divided by $d$. Dimensional forms of graphics are available in the supplementary material.}
	\label{fig:hardening_strain}
\end{figure}

Figures \ref{fig:hardening_strain}a-d show the space-time diagrams of the longitudinally averaged strain $\overline{\varepsilon}$ for $\theta / \theta_c$ from 1.4 to 3.5, respectively. We observe that, as the shear stress increases, the height of the creep layer decreases while the strain increases. In order to further investigate that, we computed the longitudinal-time averages of the strain, $\left< \overline{\varepsilon} \right>$, which we plot in Fig. \ref{fig:hardening_strain}e as a function of height $z/z_c$ for $\theta / \theta_c$ from 1.4 to 3.5. We notice two distinct regions: a region below $z/z_c$ $\approx$ 0.5, in which the levels of strain are relatively low and roughly independent of $\theta$, and a region above $z/z_c$ $\approx$ 0.5, in which the strain increases with the shear stress. We also took the maximum values of $\left< \overline{\varepsilon}\right>$, represented by $\left< \overline{\varepsilon}\right>_{max}$, which we plot in Fig. \ref{fig:hardening_strain}f. From this figure, it is possible to determine the time that each applied stress takes to cause a maximum strain of approximately $d$, which is $t_{\varepsilon}/t_{shear}$ $\sim$ 10$^6$ in dimensionless form. In dimensional terms (graphics available in the supplementary material), the characteristic time decreases with the shear stress, being $t_{\varepsilon}$ $\sim$ 10$^4$ min for $\theta / \theta_c$ = 1.4, $t_{\varepsilon}$ $\sim$ 10$^3$ min for $\theta / \theta_c$ = 2.0, $t_{\varepsilon}$ $\sim$ 10$^2$ min for $\theta / \theta_c$ = 2.7, and $t_{\varepsilon}$ $\sim$ 10 min for $\theta / \theta_c$ = 3.5. Finally, Fig. \ref{fig:hardening_strain}g shows a map in the $\theta / \theta_c$ vs. $t/t_{shear}$ space of the integrals of the $\left< \overline{\varepsilon} \right>$ profiles, $\epsilon$, divided by $d$. With this map, we can evaluate the regions in which either the fast or the slow evolution of the bed takes place for different shear stresses. Dimensional forms of the graphics are available in the supplementary material.

\begin{figure}[h!]
	\begin{center}
		\includegraphics[width=.75\linewidth]{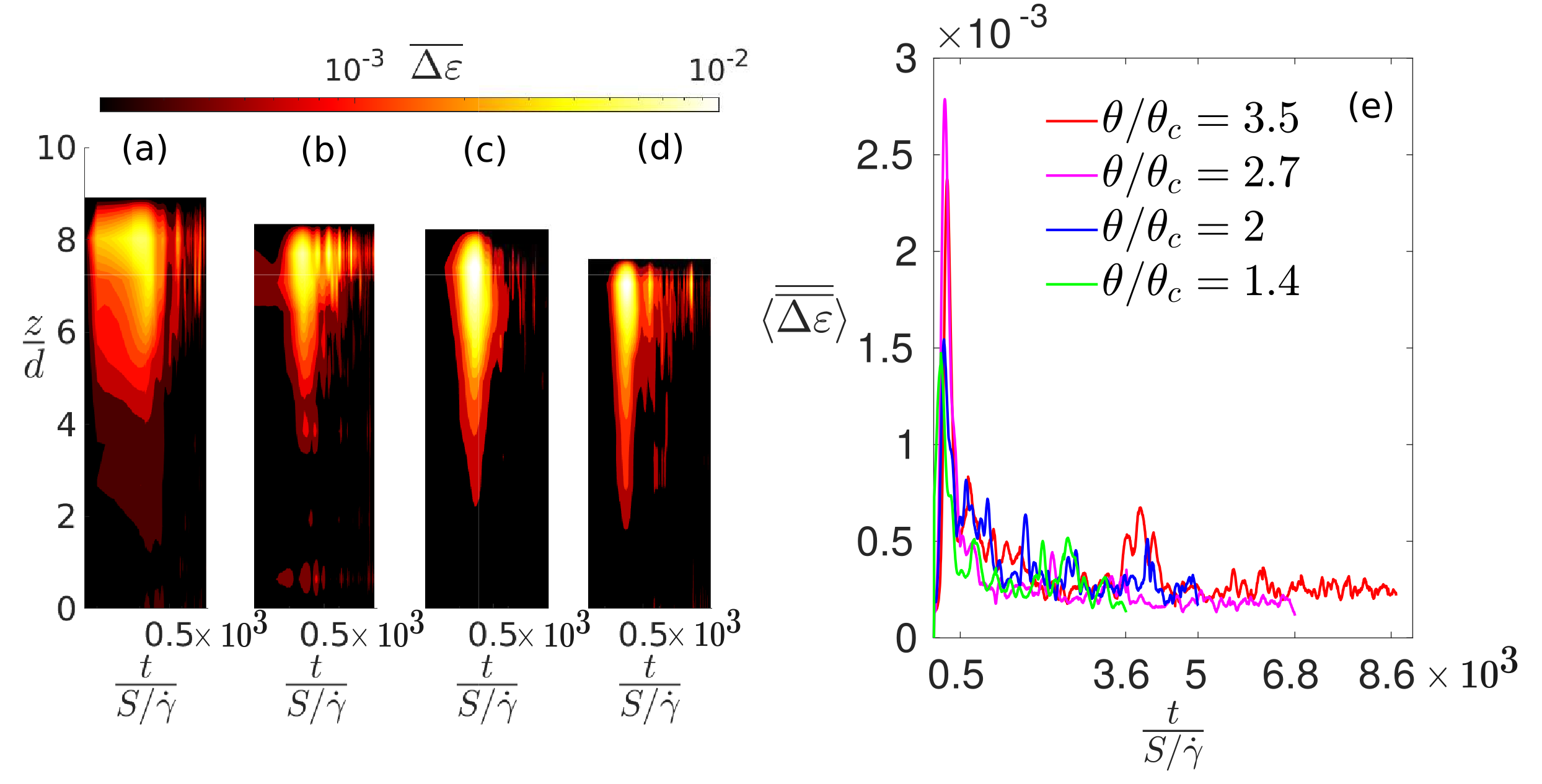}\\
	\end{center}
	\caption{(a)--(d) Space-time diagrams of the time-longitudinal average of the deformation $\overline{\Delta \varepsilon}$ for, respectively, $\theta / \theta_c$ from 1.4 to 3.5 (from left to right). (e) Longitudinal-vertical average of deformations within the creep layer, $\overline{\overline{\Delta \varepsilon}}$, as a function of time, for $\theta / \theta_c$ from 1.4 to 3.5. Graphics in dimensional form are available in the supplementary material.}
	\label{fig:hardening3}
\end{figure}

We also computed the deformation of the creep layer based on the subtraction of the previous position of each particle from its previous position, which we show in Fig. \ref{fig:hardening3}. Figure \ref{fig:hardening3}a-d shows the space-time diagrams of the longitudinally averaged deformation $\overline{\Delta \varepsilon}$ for $\theta / \theta_c$ from 1.4 to 3.5, from which we can observe that: (i) deformations are higher at the beginning of tests ($t/t_{shear}$ $\sim$ 10$^2$, in dimensional terms $t$ $\lesssim$ 1 min); (ii) as the shear stress increases, large deformations become concentrated at $t/t_{shear}$ $\approx$ 5 $\times$ 10$^2$ ($t$ $\approx$ 1 min); and (iii), as the shear stress increases, the depth reached by large deformations also increases. In this way, while for $\theta / \theta_c$ = 1.4 higher deformations are distributed within 4 $\lessapprox$ $z/d$ $\lessapprox$ 9 and $t/t_{shear}$ $\lessapprox$ 5 $\times$ 10$^2$, and for $\theta / \theta_c$ = 3.5 they occur within 2 $\lessapprox$ $z/d$ $\lessapprox$ 7 and 2 $\times$ 10$^2$ $\lessapprox$ $t/t_{shear}$ $\lessapprox$ 3 $\times$ 10$^2$. Therefore, higher shear stresses deform deeper regions in the bed during shorter times. However, the maxima in all diagrams occur at $t/t_{shear}$ $\sim$ 10$^2$ ($t$ $\approx$ 1 min). Figure \ref{fig:hardening3}e shows the time variation of the longitudinal-vertical average of deformations, $\overline{\overline{\Delta \varepsilon}}$, for $\theta / \theta_c$ from 1.4 to 3.5. We notice that $\overline{\overline{\Delta \varepsilon}}$ has a peak at the beginning of motion at $t/t_{shear}$ $=$ $t_{\Delta \varepsilon}/t_{shear}$ $\sim$ 10$^2$ ($t_{\Delta \varepsilon}$ $\sim$ 1 min) for all shear stresses tested, and that the peak tends to increase with the shear stress, corroborating the observations made for Figs. \ref{fig:hardening3}a-d.

Summarizing, we found one characteristic time for the deformation, $t_{\Delta \varepsilon}$, which is independent of the applied stress both in dimensionless and dimensional terms,

\begin{itemize}
	\item $t_{\Delta \varepsilon}/t_{shear}$ $\sim$ 10$^2$ ($t_{\Delta \varepsilon}$ $\sim$ 1 min), for any $\theta / \theta_c$,
\end{itemize}

\noindent and another one for the strain, $t_{\varepsilon}$, which in dimensional form depends on the shear stress. For strains corresponding to maximum displacements equal to $d$:

\begin{itemize}
	\item $t_{\varepsilon}/t_{shear}$ $\sim$ 10$^6$, for any $\theta / \theta_c$
	
	\begin{itemize}
	\item $t_{\varepsilon}$ $\sim$ 10$^4$ min for $\theta / \theta_c$ = 1.4;
	\item $t_{\varepsilon}$ $\sim$ 10$^3$ min for $\theta / \theta_c$ = 2.0;
	\item $t_{\varepsilon}$ $\sim$ 10$^2$ min for $\theta / \theta_c$ = 2.7;
	\item $t_{\varepsilon}$ $\sim$ 10 min for $\theta / \theta_c$ = 3.5.
	\end{itemize}

\end{itemize}

\section{CONCLUSIONS}
\label{sec:conclusions}

In this paper, we investigated the evolution of a bidisperse bed consisting of heavy grains ($S$ = 2.7) sheared by a viscous liquid. For the range of shear stresses imposed, the bed developed a bedload layer on the top of a creep layer, for which we found that: (i) there exist diffusive, advective and constrained regions for the motion of larger particles; (ii) most of segregation occurs during the very first stages of the flow (within the first 10 min, or 10$^3$ when normalized by $t_{shear}$ = $S/\dot{\gamma}$); (iii) segregation occurs within the bedload layer and in the 5\% topmost region of the creep layer; (iv) there exist four regions of increasing depth $\zeta_1$ to $\zeta_4$ where the characteristic times for segregation are $t_1/t_{shear}$ = 10$^2$, $t_2/t_{shear}$ = 10$^3$, $t_3/t_{shear}$ = 10$^4$, and $t_4/t_{shear}$ = 10$^5$--10$^6$ ($t_1$ = 1 min, $t_2$ = 10 min, $t_3$ = 10$^2$ min, and $t_4$ = 10$^3$ min). The first three regions are within the bedload layer and the last one corresponds to the top of the creep layer; (v) bed hardening becomes stronger while bedload and creep weaken along time; (vi) the characteristic time of bed hardening in terms of deformation is $t_{\Delta \varepsilon}/t_{shear}$ $\sim$ 10$^2$ ($t_{\Delta \varepsilon}$ $\sim$ 1 min), corresponding to the time when a huge peak in deformation occurs for all shear stresses; (vii) the characteristic time of bed hardening in terms of strain (corresponding to maximum values equal to $d$) is $t_{\varepsilon}/t_{shear}$ $\sim$ 10$^6$, and in dimensional form varies with the shear stress, going from $t_{\varepsilon}$ $\sim$ 10 min to $t_{\varepsilon}$ $\sim$ 10$^4$ min for $\theta / \theta_c$ decreasing from 3.5 to 1.4, respectively. Our results shed light on the complex motion of sheared beds found in nature, such as river beds and creeping lands, revealing the different layers and characteristic times for both segregation and hardening. In particular, the results can be useful for predicting the segregation in polydisperse beds (leading to bed armoring), the time for compaction of lower layers (promoting bed hardening), and the time for the rearrangement of grains within the bed (which hardens the bed while keeping memory effect \cite{Cunez2}).

\section*{AUTHOR DECLARATIONS}
\noindent \textbf{Conflict of Interest}

The authors have no conflicts to disclose

\section*{SUPPLEMENTARY MATERIAL}
See the supplementary material for  a brief description of the employed methods, the layout of the experimental setup, microscopy images of the employed grains, additional tables and graphics, and movies of sheared beds.

\section*{DATA AVAILABILITY}
The data that support the findings of this study are openly available in Mendeley Data at http://dx.doi.org/10.17632/r96kpf7ytb \cite{Supplemental}.

\begin{acknowledgments}
Erick M. Franklin and Fernando D. C\'u\~nez are grateful to the S\~ao Paulo Research Foundation -- FAPESP (Grant Nos. 2016/18189-0, 2018/14981-7) for the financial support provided. Jaime O. Gonzalez would like to thank the Petroleum Department of the Escuela Politécnica Nacional, Quito, Ecuador. The authors are also grateful to the Conselho Nacional de Desenvolvimento Cient\'ifico e Tecnol\'ogico -- CNPq (Grant No. 405512/2022-8) for the financial support provided, and to Danilo S. Borges for the help with a tracking algorithm.
\end{acknowledgments}

\bibliography{references}

\end{document}